\documentclass[twocolumn,prl,amsmath,amssymb,superscriptaddress,floatfix]{revtex4}

\usepackage{graphicx}
\usepackage{dcolumn}
\usepackage{bm}

\begin{document}

\title{Coulomb Drag as a Probe of the Nature of Compressible States in a Magnetic Field}

\author{K.~Muraki}
\affiliation{Max-Planck-Institut f\"ur Festk\"orperforschung, Heisenbergstrasse 1, 70569 Stuttgart, Germany}
\affiliation{NTT\,Basic\,Research\,Laboratories,\,NTT\,Corporation,\,3-1\,Morinosato-Wakamiya,\,Atsugi\,243-0198,\,Japan}

\author{J.~G.~S.~Lok}
\author{S.~Kraus}
\author{W.~Dietsche}
\author{K.~von~Klitzing}
\affiliation{Max-Planck-Institut f\"ur Festk\"orperforschung, Heisenbergstrasse 1, 70569 Stuttgart, Germany}

\author{D.~Schuh}
\author{M.~Bichler}
\author{W.~Wegscheider}
\altaffiliation[present address: ]{Fakult\"at f\"ur Physik, Universit\"at Regensburg, D-93040
Regensburg, Germany}
\affiliation{Walter-Schottky-Institut, Technische Universtat M\"unchen, 85748 Garching, Germany}

\date{\today}

\begin{abstract}
Magneto-drag reveals the nature of compressible states and the underlying interplay of disorder and interactions.
At $\nu=3/2$ a clear $T^{4/3}$ dependence is observed, which signifies the metallic nature of the $N=0$ Landau level.
In contrast, drag in higher Landau levels reveals an additional contribution, which anomalously grows with decreasing $T$ before turning to zero following a thermal activation law.
The anomalous drag is discussed in terms of electron-hole asymmetry arising from disorder and localization, and the crossover to normal drag at high fields as due to screening of disorder.

\end{abstract}


\maketitle

The interplay of disorder and Coulomb interactions is the key ingredient in the physics of two-dimensional electron systems (2DESs) subjected to quantizing magnetic fields.
While it is widely accepted that without interactions all states are localized by disorder except one delocalized state at the Landau-level (LL) center, in high-mobility samples electron correlations lead to a metallic  state in the lowest ($N=0$) LL\,\cite{Kalmeyer,Halperin}, and to even more exotic states in higher ($N \geq 1$) LLs in ultra-high-mobility samples\,\cite{HigherLL}.
Even in samples of moderate mobilities, Coulomb interactions are anticipated to qualitatively alter the nature of compressible states between quantized plateaus via the screening of long-range potential disorder inherent to modulation-doped structures\,\cite{Cooper}.

Coulomb drag in double-layer 2DESs\,\cite{Gramila}, in which a current driven through one layer causes a voltage in the other layer via interlayer scattering, has proven to be a powerful technique for investigating interlayer interactions.
Physically, interlayer scattering occurs via charge density fluctuations, so the drag reflects the ability of the system to create and maintain density fluctuations.
If interlayer correlation is weak or absent, drag mostly reflects the density response of individual 2DESs\,\cite{Zheng}, and thus provides information complementary to single-layer transport.
At $B=0$, drag is well understood in terms of momentum transfer between two Fermi liquids.
The phase space for scattering scales as $\sim T$ for each layer, which entails a $T^{2}$ dependence of the drag resistivity, $\rho_{D}=(W/L)\,V_{\text{drag}}/I_{\text{drive}}$ ($W/L$: width-to-length ratio of the sample).
$\rho_{D}$ is defined as positive if the voltage in the drag layer appears in the opposite direction to the resistive voltage drop in the drive layer, as is the case for drag between like-charge carriers at $B=0$.
The sign oscillations of the drag, observed in quantizing magnetic fields when the densities of the two layers are mismatched\,\cite{Feng,Lok}, are thus totally unexpected from conventional theories\,\cite{Bonsager}.
A recent theory\,\cite{von Oppen}, which considers the nonlinear dependence of conductivity on electron density, produces the required sign reversal, but in conflict with experiments, predicts negative drag for matched densities and a sizeable Hall drag component.

In this Letter, we study magneto-drag over an unprecedentedly wide range of temperatures down to 50\,mK and up to 16\,K.
The data reveal the existence of an anomalous contribution in the higher ($N \geq 1$) LLs, which is thermally activated at low $T$ and declines at higher $T$.
We discuss its origin in terms of electron-hole (e-h) asymmetry arising from disorder and localization, and the crossover to the normal drag at high fields as a manifestation of the interplay of disorder and screening.

Measurements are performed on independently contacted density-tunable double-layer 2DESs in GaAs double quantum wells (QWs)\,\cite{Lok}.
We present data for 80-$\mu$m wide Hall bars taken from two wafers that show different behavior for spin-split LLs.
Sample A\,(B) has QWs of 25\,(15)\,nm separated by a barrier of 18\,(22)\,nm.
The electron mobility of the front (back) 2DES is 140\,(80)\,m$^{2}$\!/Vs in sample A, and 120\,(105)\,m$^{2}$\!/Vs in sample B.
A drive current of 2--50\,nA at 7\,Hz is applied in one layer and the voltage in the other layer is lock-in detected.
All standard checks for drag\,\cite{Gramila} rule out spurious signals.
The samples are cooled in a $^{3}$He cryostat (0.25--16\,K) or a dilution refrigerator (0.05--0.7\,K).

\begin{figure}[!t]
\centering
\includegraphics[width=70mm]{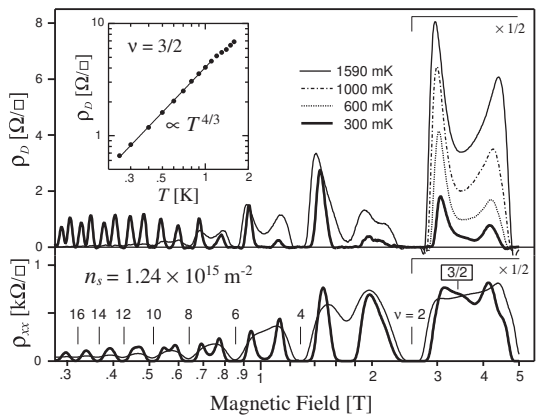}
\caption{Drag resistivity $\rho_{D}$ (top) and single-layer resistivity
$\rho_{xx}$ of the front layer (bottom) for matched densities of $n_{s}=1.24 \times 10^{15}$\,m$^{-2}$.
For visibility, $\rho_{D}$ and $\rho_{xx}$ for $\nu < 2$ are multiplied by 0.5.
Inset: $T$ dependence of $\rho_{D}$ at $\nu=3/2$, demonstrating the $T^{4/3}$ power law.}
\label{Bdep}
\end{figure}

We first show data of sample A.
Figure\,\ref{Bdep} plots typical traces of $\rho_{D}$ (upper panel) and the single-layer resistivity $\rho_{xx}$ of the front 2DES (lower panel) for matched densities of $n_{s}=1.24\times 10^{15}$\,m$^{-2}$.
As in earlier experiments\,\cite{Rubel,Hill}, $\rho_{D}$ oscillates with magnetic field $B$ reflecting the density of states (DOS) at the Fermi level ($E_{F}$) and hence the phase space for interlayer scattering.
The data additionally reveal two important points.
First, the minima in $\rho_{D}$ at odd integer fillings persist down to very low fields where $\rho_{xx}$ does not show any spin splitting.
The origin of these minima, initially discussed in terms of higher sensitivity of drag to spin splitting\,\cite{Hill} or enhanced screening of interlayer interactions at the DOS maxima\,\cite{Rubel}, was later put into question\,\cite{Lok}, and will be explained below in terms of e-h symmetry.
Second, the temperature dependence of $\rho_{D}$ is different for high and low fields.
We find that at LL filling factor $\nu=3/2$ the drag follows a power law, $\rho_{D}\propto T^{4/3}$, below 1.2\,K (inset).
($\nu=n_{s}/n_{\phi}$ with $n_{\phi} \equiv eB/h$ the degeneracy of a spin-split LL.)
The power dependence reflects the continuous DOS at $E_{F}$ and hence gapless excitations available for interlayer scattering, thus signifying the metallic nature of the $N=0$ LL at half filling, similar to the case for $\nu = 1/2$\,\cite{Lilly}.
The observation of the $T^{4/3}$ dependence, in contrast to $T^{2}$ for simple Fermi liquids, agrees with theoretical predictions for drag between composite fermions (CF)\,\cite{CompositeFermion}, thus providing evidence for the predicted  unusual frequency and wave vector dependence of the conductivity\,\cite{Halperin}.

\begin{figure}[!t]
\centering
\includegraphics[width=80mm]{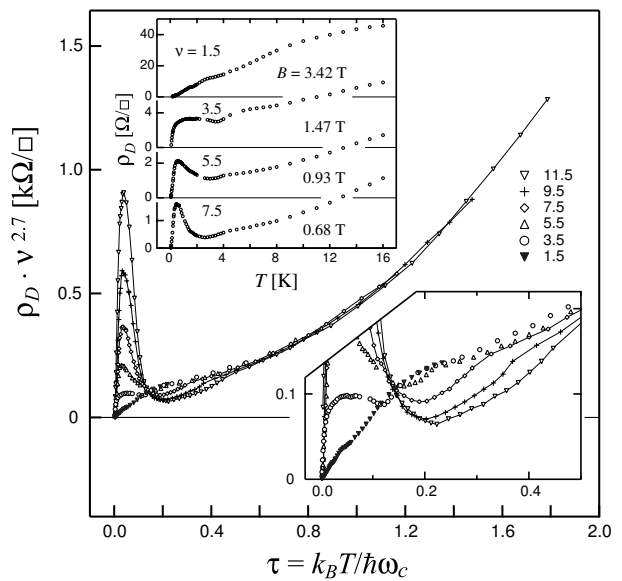}
\caption{Temperature dependence of $\rho_{D}$ at various half-integer fillings.
Left inset: $\rho_{D}$ for $\nu=1.5$--7.5.
Main panel: data for $\nu=1.5$--11.5 plotted against normalized temperature, $\tau=k_{B}T/\hbar \omega_{c}$.
To scale the data, the vertical axis is taken to be $\rho_{D}\,\nu^{2.7}$.
Right inset: magnification of the low-$T$ section of the plot in the main panel.
}
\label{Tdep}
\end{figure}

Turning to the higher ($N \geq 1$) LLs, we note that the $T$ dependence gets weaker with decreasing $B$, and eventually becomes opposite, i.e., $\rho_{D}$ is increasing with decreasing $T$.
The left inset of Fig.\,\ref{Tdep} compares the $T$ dependence over a much wider temperature range for various fillings, $\nu=2N + 3/2$ ($N=0,1,2,3$), which for complete spin splitting would correspond to the half-filled spin-down branch of the $N^{\text{th}}$ LL.
The data clearly reveal the existence of two contributions in the higher LLs.
In addition to the {\it normal} drag which monotonically increases with $T$ and dominates at high $T$, there is an {\it anomalous} contribution at low $T$ which initially grows with $T$ and then turns to decrease upon further increasing $T$.
A decrease in $\rho_{D}$ with increasing $T$ is totally inconsistent with what is expected from the simple phase-space argument.
We further note that the anomalous drag eventually vanishes for $T \rightarrow 0$, indicating that it originates from thermal fluctuations rather than interlayer correlations\,\cite{Kellogg}.

To elucidate the contributions of normal and anomalous drag for different $\nu$, Fig.\,\ref{Tdep} replots the data against normalized temperature, $\tau \equiv k_{B}T/\hbar \omega_{c}$, where $\omega_{c}=eB/m^{\ast}$ is the cyclotron frequency and $m^{\ast}= 0.067 m_{0}$ the effective mass.
To accommodate the factor $B^{-1}$ in the abscissa and the $B$ dependence of the drag, the vertical axis is taken to be $\rho_{D}\,\nu^{\alpha}$.
Here we choose $\alpha=2.7$, which conveniently scales the high-$T$ data for all $\nu$.
The use of the normalized temperature is necessary since the normal drag for different $\nu$ cannot be described by one single power law.
An approximate $\tau^{2}$ dependence is restored only in the high-$T$ regime ($\tau \gtrsim 1.0$), where inter-LL scattering is the dominant source of drag.
For $\tau \lesssim 0.2$, where intra-LL excitation is mostly relevant, the exponent is smaller, but cannot be quantified for the higher ($N \geq 1$) LLs due to the presence of the anomalous contribution which develops at these temperatures

\begin{figure}[!t]
\centering
\includegraphics[width=72mm]{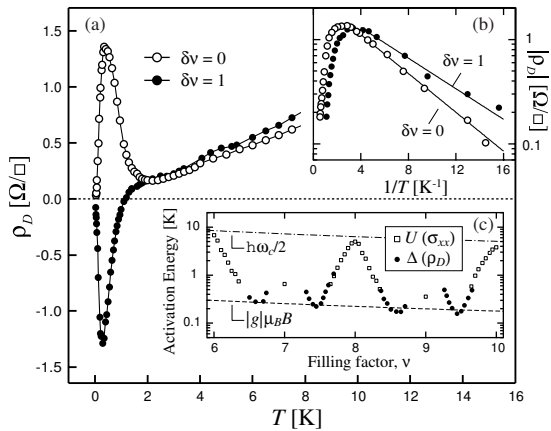}
\caption{(a) $\rho_{D}$ for matched and mismatched densities with $(\nu_{F},\nu_{B})=(9.5,9.5)$ and $(8.5,9.5)$.
(b) Arrhenius plot of $\lvert \rho_{D} \rvert$ at low $T$, demonstrating the activated behavior.
(c) measured activation energies, $\Delta$ ($U$), of $\rho_{D}$ ($\sigma_{xx}$) for matched densities vs.\ $\nu$.
Data of $U$ are shown for the front layer only, in regions where $\sigma_{xx} \propto \exp (-U/T)$.
The bare Zeeman energy in GaAs with $\vert g \vert = 0.4$ is shown as a reference.
}
\label{Mismatch}
\end{figure}

Another distinct feature of the anomalous drag is the bipolarity of the sign, and its dependence on the filling-factor difference, $\delta \nu$ ($\equiv \nu_{B}-\nu_{F}$), between the layers.
This is demonstrated in Fig.\,\ref{Mismatch}(a), which compares $\rho_{D}$ for the filling-factor combinations of $(\nu_{F},\nu_{B})=(9.5,9.5)$ and $(8.5,9.5)$ in the front and back layers.
For $T \gtrsim 2$\,K where normal drag dominates, there is no qualitative difference, while at low $T$ where anomalous drag dominates, $\rho_{D}$ turns negative for $\delta \nu =1$, but with the absolute value showing similar behavior as the positive drag for $\delta \nu =0$.
This clearly indicates that the positive drag for $\delta \nu =$ even and negative drag for $\delta \nu =$ odd have the same origin.

Furthermore, instead of the generally assumed power dependence, we find that the low-$T$ behavior of the anomalous drag is well described by thermal activation, $\vert \rho_{D} \vert \propto \exp (-\Delta/T)$ [Fig.\,\ref{Mismatch}(b)].
This thermal activation indicates the existence of a finite energy gap for creating density fluctuations, thus suggesting a different nature of states in higher LLs.
Figure\,\ref{Mismatch}(c) compares the activation energy $\Delta$ of $\rho_{D}$ for matched densities with the activation energy $U$ of the single-layer conductivity $\sigma_{xx}$ calculated from $\rho_{xx}$ and $\rho_{xy}$.
The two energies depend on $\nu$ similarly, reflecting the position of $E_{F}$ with respect to the delocalized level at the LL center, and in the regions where they overlap, their magnitudes are comparable.
These observations clearly indicate the relevance of localized states in the anomalous drag\,\cite{Localization,Mortensen}.
We note, however, that $\Delta$ remains finite even at the spin-split conductivity peaks where $\sigma_{xx}$ does not show activated behavior.

Upon further discussing the origin of the anomalous drag, we recall that finite drag results from different propagation of electronlike (occupied states above $E_{F}$) and holelike (empty states below $E_{F}$) excitations within an energy window $\sim k_{B}T$ around $E_{F}$, which is often referred to as `e-h asymmetry', or `nonlinear susceptibility'\,\cite{Kamenev,von Oppen}.
For normal Fermi liquids, this is related to the curvature of the energy dispersion, $\partial^{2} E/\partial k^{2}$, which determines if an electron excited above $E_{F}$ has a larger velocity than the `hole' left below $E_{F}$.
Such arguments, however, do not hold for quantum Hall systems, where in the absence of disorder LLs have no dispersion in the interior of the sample.
Only in the $N=0$ LL does a Fermi surface form as a result of strong electron correlations.
In Ref.\,\cite{Feng}, negative drag has been discussed in terms of holelike dispersions, $\partial^{2} E / \partial k^{2} < 0$, of bulk extended states or edge states caused by disorder.
Consideration of only continuum of extended states, however, would lead to power dependence, rather than the observed activated behavior.
Reference\,\cite{von Oppen}, on the other hand, proposed that the source of the nonlinear susceptibility is the dependence of the local conductivity on the local electron density.
The drag and its sign would thus reflect the product of ${\rm d} \sigma_{xx}/ {\rm d} n_{s}$ of the two layers, which does explain why the anomalous drag vanishes at the {\em spin-degenerate} $\sigma_{xx}$ peaks (Fig.\,\ref{Bdep}, $\nu=11, 13, 15, 17$), while in seeming conflict with the data for the {\em spin-split} case where $\rho_{D}$ does not show minima at the $\sigma_{xx}$ peaks.

\begin{figure}[!t]
\centering
\includegraphics[width=75mm]{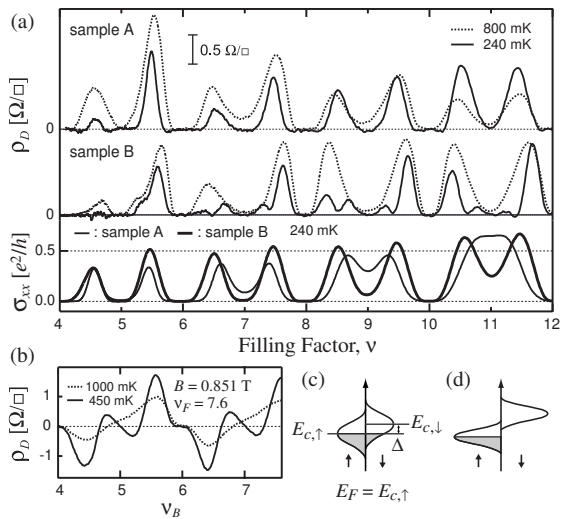}
\caption{(a) $\rho_{D}$ (upper) and single-layer conductivity $\sigma_{xx}$ (lower) as a function of $\nu$ for matched densities [$n_{s}=1.24$ (1.57) $\times 10^{15}$\, m$^{-2}$ in sample A (B)].
(b) $\rho_{D}$ vs.\ $\nu_{B}$ of sample B for fixed $\nu_{F}$ ($=7.6$).
(c), (d) Illustrations of the populations of up and down spin states in the topmost LL at the $\sigma_{xx}$ maxima in (c) sample A and (d) sample B.}
\label{SampleB}
\end{figure}

To settle this enigma, we note that due to broadening of LLs by disorder, electrons of both spin directions can coexist at $E_{F}$, and that the energies of the delocalized states for the spin-up and down electrons, $E_{c,\uparrow}$ and $E_{c,\downarrow}$, are separated by the (exchange enhanced) Zeeman energy [Fig.\,\ref{SampleB}(c)].
When $E_{F}=E_{c,\uparrow}$, the spin-up electrons in the extended state dominate $\sigma_{xx}$.
This spin-up branch at its half filling, however, possesses e-h symmetry, and would not contribute to the anomalous drag.
Note however, that the e-h symmetry of the overall system is broken by the population of spin-down branch below its half filling.
In this picture, the activation energy corresponds to the energy difference between $E_{F}$ and the delocalized state at the LL center of the {\em minority} spin.
At half filling of the majority spin, this corresponds to the Zeeman energy [Fig.\,\ref{SampleB}(c)], and upon moving away from half filling, it increases\,\cite{Activation}, consistent with the experiment [Fig.\,\ref{Mismatch}(c)].

The above arguments are confirmed by the results of sample B, shown in Fig.\,\ref{SampleB}(a) for matched densities of $n_{s}=1.57\times 10^{15}$\,m$^{-2}$.
For comparison, data of sample A are included.
While at 800\,mK the two samples behave similarly, at lower $T$ new features develop.
$\rho_{D}$ of sample B now shows additional minima near the maxima of spin-split $\sigma_{xx}$ peaks, most clearly seen at $\nu=8.5$--10.5.
When the filling factor in one layer is varied, $\rho_{D}$ additionally changes sign around these fillings [Fig.\,\ref{SampleB}(b)].
These observations indicate that in sample B the system acquires e-h symmetry at the maxima of the spin-split $\sigma_{xx}$ peaks, suggesting that the electrons in the topmost LL are fully spin polarized as schematically shown in Fig.\,\ref{SampleB}(d).
This is also supported by the $\sigma_{xx}$ data.
In sample A, spin-up and spin-down $\sigma_{xx}$ peaks are overlapping, and accordingly, $\sigma_{xx}$ peaks are displaced toward odd-integer fillings.
By contrast, $\sigma_{xx}$ peaks of sample B appear closer to half fillings, indicating a more complete spin polarization\,\cite{Fogler}.
Why the two samples having similar mobilities exhibit such different behavior is not clear.
One interesting possibility is that the narrower well widths of sample B enhance the {\em intralayer} exchange interactions, which facilitates full spin polarization in the topmost LL.

Now we discuss how the involvement of localization modifies the e-h (a)symmetry argument.
Following the line of arguments in Ref.\,\cite{von Oppen}, we speculate that the e-h asymmetry underlying the anomalous drag is associated with the response of localization properties to local potential modulations.
For instance, lowering (raising) the local potential at a saddle point connects the adjacent closed orbits trapped around potential valleys (hills)\,\cite{Polyakov}.
The involvement of localized states could thus explain why in general (as in sample A) $\sigma_{xx}$ and $\rho_{D}$ do not show the same e-h symmetry, i.e., the former probes only extended states while the latter probes localized states as well.
Furthermore, why the anomalous drag declines at high $T$ is naturally explained.
Once the thermal distribution gets wider than the amplitude of the disorder potential, the sensitivity of the localization/delocalization to local potential modulations is lost, and so is the anomalous drag.
The use of a similar idea as in Ref.\,\cite{von Oppen}, however, would result in  the overall sign of $\rho_{D}$ opposite to the experimental results.
Further advances in theories are required to account for the overall sign,  huge magnitude, and temperature dependence of the anomalous drag.

Finally, we address how the anomalous drag is taken over by the normal one with increasing $B$, or why negative drag is observed only at low $B$\,\cite{Feng,Lok_Physica}.
At low $B$, due to the small DOS ($n_{0}$) of the partially occupied LL, the disorder potential is not fully screened.
As a result, electrons in the uppermost LL split into incompressible regions that are either locally full or empty.
The resultant potential landscape is similar to that of a noninteracting 2DES.
As the DOS per LL gets larger with increasing $B$, compressible regions, in which the disorder potential is fully screened, start to develop, which for the $N=0$ LL evolve into the CF liquid.
Note that the e-h picture underlying the anomalous drag is no longer valid for the latter case where the potential becomes flat.
At $T=0$, the crossover between the two regimes occurs when the scale of the Coulomb force, $n_{0}\,e^{2}\!/2\pi \epsilon$, exceeds the root mean square of the bare potential gradient\,\cite{Cooper}, $\sqrt{ \left\langle \vert \nabla W_{B}\vert^{2}\right\rangle} = \sqrt{n_{D}/32\pi}\,e^{2}\!/\epsilon w$, with $\epsilon$ the dielectric constant, $n_{D}$ the density of remote ionized donors, and $w$ the spacer thickness.
With $w=60$\,nm and $n_{D}=n_{s}$, the crossover occurs around 0.76 and 1.52\,T for the spin-degenerate ($n_{0}=2n_{\phi}$) and fully spin-split ($n_{0}=n_{\phi}$) cases, respectively.
These values are consistent with the typical maximum field ($\sim 1$\, T) at which negative drag is observed in high-mobility samples\,\cite{LowMobility}.

In summary, anomalous drag in higher LL at low fields suggests the relevance of localized states.
The crossover to the normal drag at high fields reflects the development of compressible regions due to increased screening of disorder, and the absence of anomalous drag in the lowest LL signifies the metallic nature specific to the $N=0$ LL.

We thank A.~H.~MacDonald, A.~Stern, F.~von Oppen, K.~Flensberg, W.~Metzner, S.~Brener, and Y.~Tokura for discussions, J.~Weis for the use of his experimental facilities, H.~Yamaguchi and Y.~Hirayama for supporting the research.
The work was supported by NEDO international joint research, `nanoelasticity' and BMBF.



\begin{references}
\bibitem{Activation}
The increase of $\Delta$ upon moving toward odd $\nu$ is understood as due to exchange enhancement of spin splitting.
\end{references}
\end{document}